%% file: BERT2Code.tex
\documentclass[runningheads]{llncs}
\usepackage{graphicx}
\usepackage{xcolor}
\usepackage{amsfonts}
\usepackage{multirow}
\usepackage{ifsym}
\usepackage{hyperref}

\makeatletter
\newcommand{\printfnsymbol}[1]{%
  \textsuperscript{\@fnsymbol{#1}}%
}
\makeatother

\begin{document}
\title{BERT2Code: Can Pretrained Language Models be Leveraged for Code Search?}
\titlerunning{BERT2Code}

%there should be change in the author order
% failed to insert equal contributions/corresponding author footnote
\author{
% First Author\inst{1}\orcidID{0000-1111-2222-3333} \and
% Second Author\inst{1}\orcidID{1111-2222-3333-4444} \and
% Third Author\inst{1}\orcidID{2222--3333-4444-5555}
Abdullah Al Ishtiaq\thanks{These authors contributed equally to this work.} \and 
Masum Hasan\printfnsymbol{1} \and
Md. Mahim Anjum Haque \and 
\\Kazi Sajeed Mehrab \and
Tanveer Muttaqueen \and 
Tahmid Hasan \and 
\\Anindya Iqbal \and
Rifat Shahriyar
}
\authorrunning{Abdullah Al Ishtiaq et al.}
\institute{Bangladesh University of Engineering and Technology, Dhaka, Bangladesh\\
\email{1505080.aai@ugrad.cse.buet.ac.bd, masum@ra.cse.buet.ac.bd, \{mahimanzum, ksmehrab, tanveer.mutta\}@gmail.com, \{tahmidhasan, anindya, rifat\}@cse.buet.ac.bd}
}

\maketitle              % typeset the header of the contribution
%

\input{Sections/Abstract}
\input{Sections/Introduction}
\input{Sections/Background}

\input{Sections/Methodology}
\input{Sections/Results}

\input{Sections/Related_Works}
\input{Sections/Conclusion}
%\input{Sections/Acknowledgments}

\bibliographystyle{splncs04}
\bibliography{references}

\end{document}

%% file: Sections/Abstract.tex
% \begin{abstract}
Millions of repetitive code snippets are submitted to code repositories every day. To search from these large codebases using simple natural language queries would allow programmers to ideate, prototype, and develop easier and faster. Although the existing methods have shown good performance in searching codes when the natural language description contains keywords from the code \cite{husain2019codesearchnet}, they are still far behind in searching codes based on the semantic meaning of the natural language query and semantic structure of the code. In recent years, both natural language and programming language research communities have created techniques to embed them in vector spaces. In this work, we leverage the efficacy of these embedding models using a simple, lightweight 2-layer neural network in the task of semantic code search. We show that our model learns the inherent relationship between the embedding spaces and further probes into the scope of improvement by empirically analyzing the embedding methods. In this analysis, we show that the quality of the code embedding model is the bottleneck for our model's performance, and discuss future directions of study in this area.

%% file: Sections/Introduction.tex
\section{Introduction}
\label{sec:bert2code-introduction}

Since the inception of computers, researchers have been dreaming of programming them with natural language instructions only \cite{imitation-game}. Although this problem is far from solved, a subset of this problem, `Semantic Code Search', has gained overwhelming traction in recent years \cite{unif,feng2020codebert,gu2018deep,gu2016deep,husain2019codesearchnet,sachdev-ncs,yao2019coacor}.

Semantic code search refers to searching for a source code with a natural language query by utilizing the inherent meaning of both the source code and the query. It has a significant impact on a wide range of computer science applications. For example, searching for source code on websites like Stack Overflow is an integral part of modern software development \cite{query-search,query-search2}. Easily finding the relevant code snippet can remarkably reduce time, effort, and project cost. Thus, for decades, researchers have been trying to search source codes automatically \cite{1994-paper}. 

With the rapid advancement of deep learning in recent years, there have been many attempts to use neural network based code search methods \cite{unif,feng2020codebert,gu2018deep,husain2019codesearchnet}. However, these methods often fail to capture the semantic meanings of the query and the source code, and instead heavily rely on common tokens between the two \cite{unif,husain2019codesearchnet,sachdev-ncs}. 

In recent years, there has been tremendous advancement in embedding models of natural language and programming language. They have shown notable performance in capturing the semantic meanings of sentences and the source codes in embedding spaces \cite{alon2019code2vec,devlin2018bert,feng2020codebert,reimers2019sbert}. In addition, translating from one sentence embedding space to another has shown promising results in Neural Machine Translation \cite{balm}. We have incorporated this idea to semantic code search by leveraging the efficacy of embedding models. In this work, we propose BERT2Code -- a simple neural network based code search method that utilizes the generalization capabilities of the state-of-the-art pretrained natural language and source code embedding models, namely Sentence-BERT \cite{reimers2019sbert}, Code2Vec \cite{alon2019code2vec}, and CodeBERT \cite{feng2020codebert}. We show that our method can capture the inherent relationship between the two embedding spaces to a reasonable extent and achieve 15.478\% MRR in code search. With manual and statistical analysis of the embedding models, we find that leveraging the full potential of the embeddings requires code embedding models with better quality and more generalization capability.

We organize our paper by first defining relevant techniques and terminologies used throughout our paper (Section \ref{sec:background}). In the section that follows, we describe our model in detail (Section \ref{sec:bert2code-methodology}). In the next section, we outline the results and findings from our study (Section \ref{sec:bert2code-results}). We conclude the paper by discussing related works from the literature (Section \ref{sec:bert2code-related-work}) and with a brief summary of our work in the conclusion (Section \ref{sec:bert2code-conclusion}).

%% file: Sections/Background.tex
\section{Background}
\label{sec:background}
In this section, we describe the necessary concepts and technologies relevant to our work.

\subsection{Sentence Embedding}
\label{subsec:background-nl-embedding}
Sentence Embedding is referred to as the vector representations of sentences. Several classical and neural approaches have been proposed for generating sentence embeddings over the years. \cite{laser,devlin2018bert,joulin2016fasttext,doc2vec,Mikolov_word2vec,reimers2019sbert,tf-idf}.

\textbf{B}idirectional \textbf{E}ncoder \textbf{R}epresentations from \textbf{T}ransformers (\textbf{BERT}) \cite{devlin2018bert} is a Transformer \cite{vaswani2017attention} based language model that is trained for two tasks of predicting masked tokens and predicting whether two given sentences are consecutive sentences or not. 

These cleverly designed tasks allow the model to be trained on large volumes of easily accessible unlabeled data from the internet, and consequently, with sufficient amount of data, the model learns to create rich semantic latent representations of the input texts. With finetuning, these representations can be used to achieve state-of-the-art performance and even above human-level performance \cite{devlin2018bert} in numerous text analysis tasks.

Based on the latent text representations created by BERT, Reimers et al. \cite{reimers2019sbert} has proposed a sentence embedding method named \textbf{Sentence-BERT} or \textbf{SBERT}, by performing a pooling operation on the output of BERT and fine-tuning BERT using siamese and triplet network structures. This resulted in updated sentence embeddings that are more meaningful and can be compared using cosine-similarity. Regardless of its name, this method is also capable of embedding multiple sentences.

\subsection{Code Embedding}
\label{subsec:background-code-embedding}
The idea of generating embeddings or vectors from source code has showed encouraging results recently. This is largely due to the success of deep learning based latent representations of source codes.  

\textbf{Code2Vec} \cite{alon2019code2vec} is a major contribution in the area of code embedding models. It represents code snippets as continuous distributed vectors or code embeddings. A given code snippet is decomposed into its abstract syntax tree and the paths in the tree are represented by vectors which are then aggregated into a single embedding using the attention mechanism \cite{bahdanau2014neural}. The authors trained their model with 12 million Java methods for the end task of predicting corresponding method names. Their model is able to generalize further than the original target and even predict method names that are unobserved in the training data. 

\textbf{CodeBERT} \cite{feng2020codebert} is another recent contribution in the concerned area. It uses exactly the same architecture as RoBERTa-base \cite{roberta} along with multi-layer bidirectional Transformer  \cite{vaswani2017attention}, and is pretrained using both unimodal and bimodal data from the CodeSearchNet Challenge dataset \cite{husain2019codesearchnet}. It has shown to outperform the baseline models of the challenge, and improves the performance in code document generation task as well.

\subsection{Similarity Search}
\label{subsec:similarity_search}
Similarity search is the problem of finding a set of objects that are the most similar to a given object. For embeddings or vectors, the problem boils down to finding vectors from a given dataset that are closest to the query-vector in terms of Euclidean or cosine distance. Similarity search using vectors works particularly well because the vector representation of objects is designed to produce similar vectors for similar objects \cite{vector-similar-1,doc2vec}.

\label{subsec:background_faiss}
The FAISS\footnote{\url{https://github.com/facebookresearch/faiss}} library, developed by Facebook AI Research, can efficiently perform similarity search and clustering on embeddings or vectors. This work has proven to largely reduce the time required for searching the nearest neighbors of a vector in a high dimensional embedding space when GPU is available \cite{faiss}. In addition, this library is highly scalable and, therefore, one can easily perform a $k$-nearest-neighbor search within a reasonable amount of time even in a dataset of size in billions.

%% file: Sections/Methodology.tex
\section{Methodology}
\label{sec:bert2code-methodology}
In this section, we describe the dataset used, the techniques to create the embeddings of natural language and code, the architecture of the neural network, and the process of training and evaluating the network. 

\subsection{Problem Definition}
\label{sec:method_problem_defn}
Given a natural language (NL) query, and a code snippet corpus, the task is to find the code snippet that best matches the NL query in terms of semantic meanings.

For a programming language $L_P$, we consider a $K_1$ dimensional embedding space $S_P$, a subset of $I\!R^{K_1}$ , where every point in the vector space represents a program $P$ written in language $L_P$. $V_P$ is an embedding of a program $P$, and $V_P$ is referred to as a code vector. Program $P$ is also referred to as a source code.
Similarly, for natural language $L_N$, we consider a $K_2$ dimensional embedding space $S_N$, a subset of $I\!R^{K_2}$, where every point in the vector space represents a description $N$ of a program in natural language $L_N$. $V_N$ is an embedding of a natural language statement $N$, and $V_N$ is referred to as an NL vector.

Given the embedding spaces $S_P$ and $S_N$, we aim to find a transformation $T$ that maps every vector in space $S_N$ to its programming language equivalent to $S_P$ with a very small number of data samples in $S_N$ and $S_P$. Given a natural language description $N$, we find its vector representation $V_N$ and by applying transformation $T$, we obtain a code vector $\hat{V}_P$. After that, we find its closest $n$ code embeddings $\{V_{P_1}, ..., V_{P_n}\}$ from a predefined set of code vectors  where $n$ $\epsilon$ $\mathbb{N}$ is an arbitrary constant. Finally, we return the codes $\{P_1, ..., P_n\}$ associated with the code vectors.

\subsection{Dataset}
\label{dataset}
In this study, we build a model to map a natural language query to its programming language counterpart. Hence, our task requires a dataset where each datapoint consists of a source code and a Natural Language (NL) query that describes the functionality of the source code in adequate detail. Upon exploring the literature, we find the CodeSearchNet Challenge \cite{husain2019codesearchnet} dataset to be of the most suitable for our use-case.
The dataset contains in total 2.3 million code-description pairs from 6 different programming languages. However, in our study, we only limit our scope to 543k given samples from Java programming language.  We use original train-valid-test split from the dataset. The detailed process of procurement and cleaning of the dataset is described in the original article \cite{husain2019codesearchnet}.

\begin{figure}[h]
    \centering
    \includegraphics[width=\linewidth]{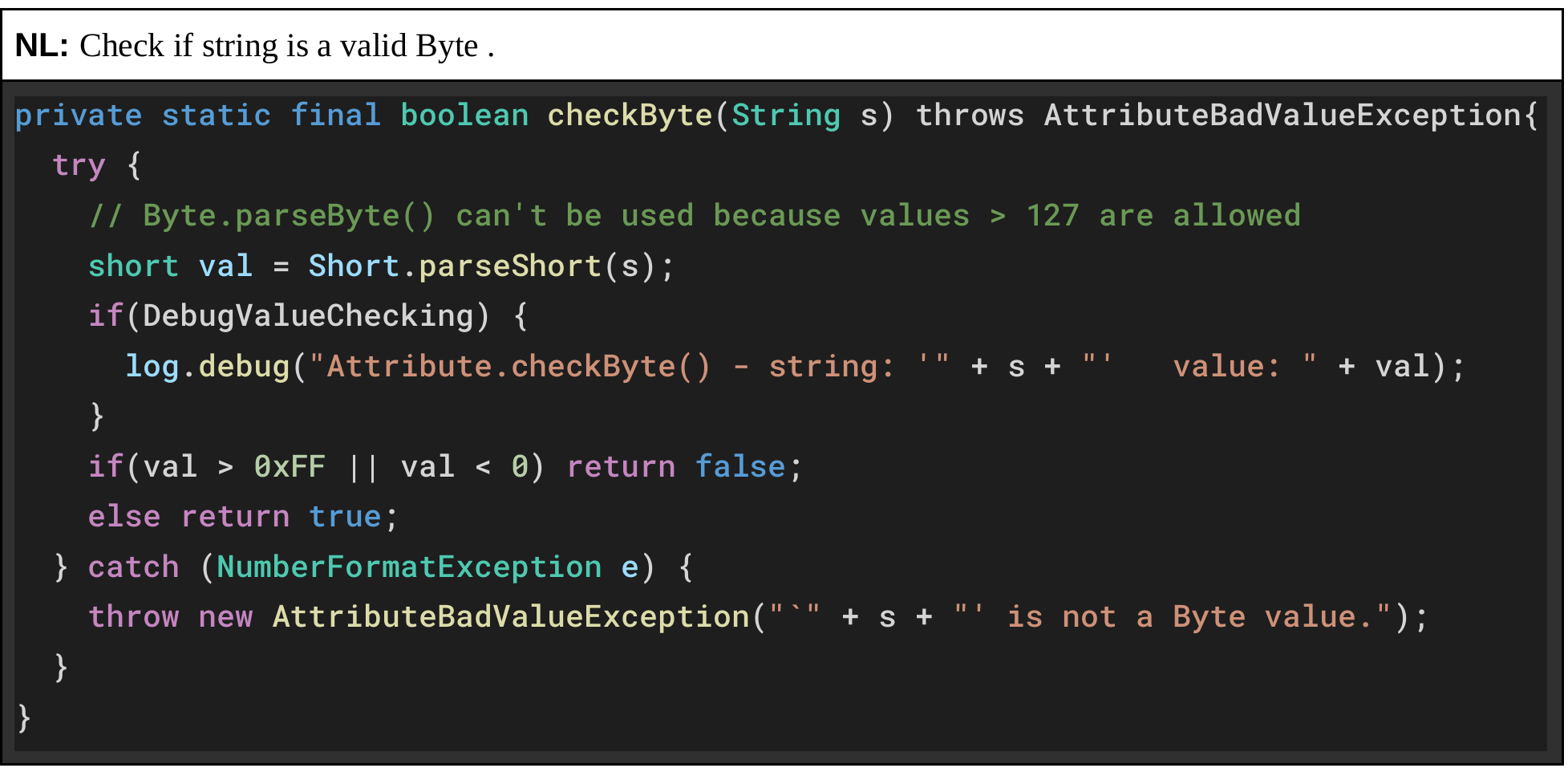}
  % \vspace{-7mm}
    \caption{Sample natural language description and source code pair from the CodeSearchNet \cite{husain2019codesearchnet} dataset}
    \label{fig:naps}
\end{figure}

\subsection{Generating Sentence Embeddings}
\label{sec:method_nl_embeddings}

In order to generate sentence embeddings, we have used Sentence-BERT (SBERT) \cite{reimers2019sbert}. Our used model employs the BERT\textsubscript{LARGE} model, uses the MEAN pooling strategy on the output, and is finetuned with Natural Language Inference (NLI) datasets \cite{nli1,nli2}  using a 3-way softmax classiﬁer objective function. We use this particular model because it shows the greatest average accuracy in generating meaningful sentence embeddings \cite{reimers2019sbert}.
Creating the NL vectors using pre-trained SBERT has taken 4.4s on an average per 1000 samples in a workstation with Intel Core i5-9600k 3.70 GHz CPU and NVIDIA GeForce RTX 2070 SUPER GPU. All of the following computations are also done in the same workstation.

\subsection{Generating Code Embeddings}
\label{sec:method_code_embeddings}
At first, we use Code2Vec \cite{alon2019code2vec} to generate code embeddings from our data. Each of the raw Java methods from the dataset is passed through the Code2Vec preprocessor. Then, we pass the preprocessed source programs through the pretrained Code2Vec model and generate the corresponding code vectors. Generating these code vectors has taken 2030s on an average per 1000 samples.

In addition to Code2Vec, we have generated code vectors using the CodeBERT \cite{feng2020codebert} model which can be used in other downstream tasks. Generating the code vectors in this case has taken 44.25s on an average per 1000 samples.   

\subsection{Training a Neural Network to Transform NL Vectors to Code Vectors}
\label{sec:training}

\begin{figure}
\centering
\includegraphics[width=.7\linewidth]{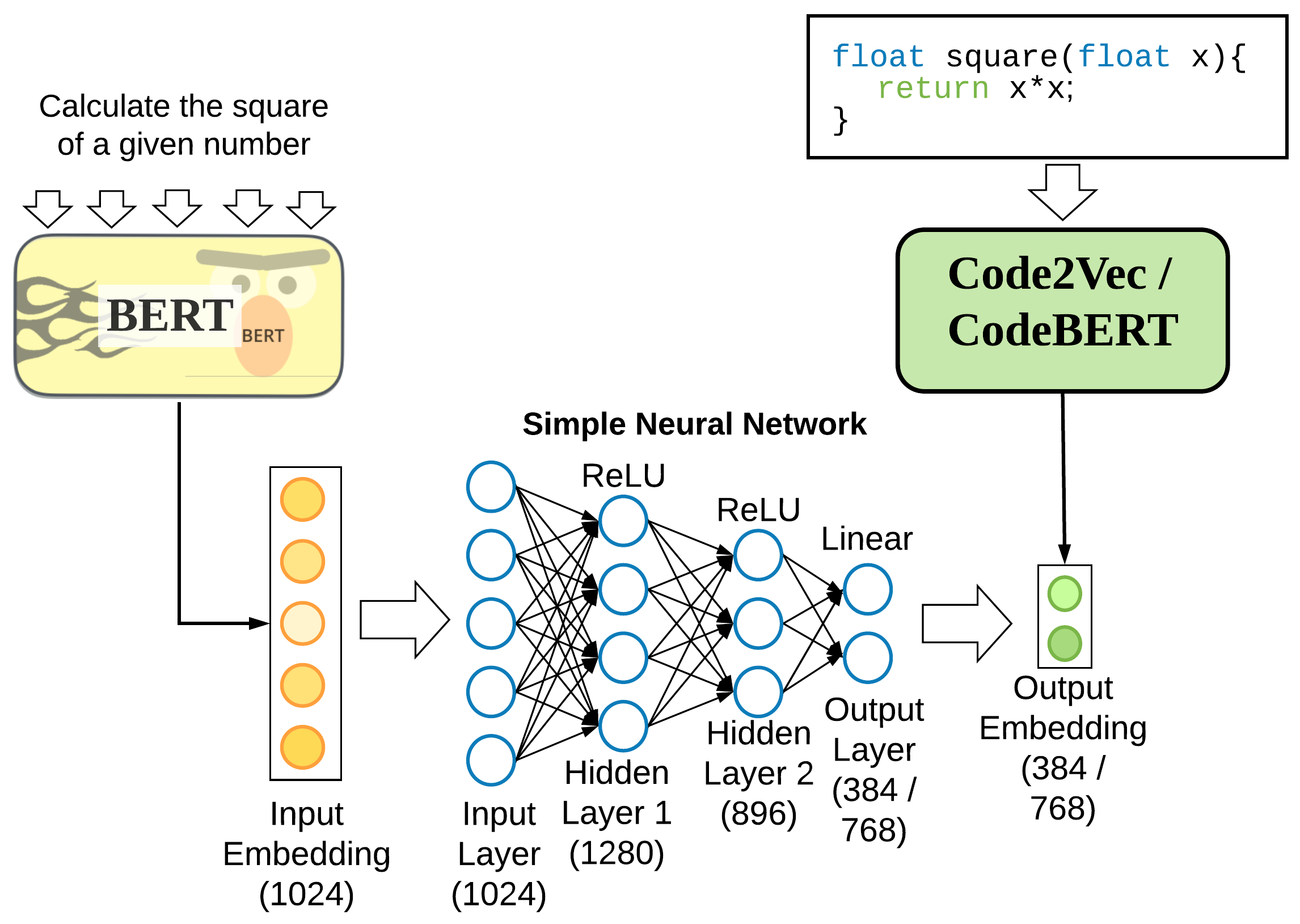}
\caption{Training a feed-forward neural network to transform natural language embedding from BERT to code embedding (BERT figure \cite{illustrated-bert})}
\label{fig:train}
\end{figure}

Once the natural language vectors  and their corresponding code vectors are produced, we train a feed-forward neural network \cite{nn} as the transformation function $T$ between an NL vector and a code vector. The neural network has 2 hidden layers with size 1280 and 896 respectively. While the size of the input layer is 1024, and the size of the output layer is 384 in case of Code2Vec and 768 in case of CodeBERT. Both of the hidden layers use ReLU \cite{relu-2,relu}, a non-linear activation function, and the output layer uses linear activation. As the loss function for our network, we use Euclidean distance with max-margin approach. 
In other words, over the training period, the network tries to reduce the L2 distance between the network prediction and the original code vector while, simultaneously, trying to increase distances from other datapoints in the code embedding space. 
We use constant learning rate of $10^{-5}$ and the batch size is kept at 16. 
We have also used early stopping mechanism \cite{kim-conv-early-stop} in order to prevent over-fitting. The training continued for 207 and 170 epochs at 5h 37m 40s and 4h 22m 32s for the two networks respectively. The network is implemented with the PyTorch Framework \cite{pytorch}. The training procedure is visualized in Figure \ref{fig:train}. 

\vspace*{-1cm}

\subsection{Predicting Code for Unseen Query}
\begin{figure}
\centering
\includegraphics[width=0.7\linewidth]{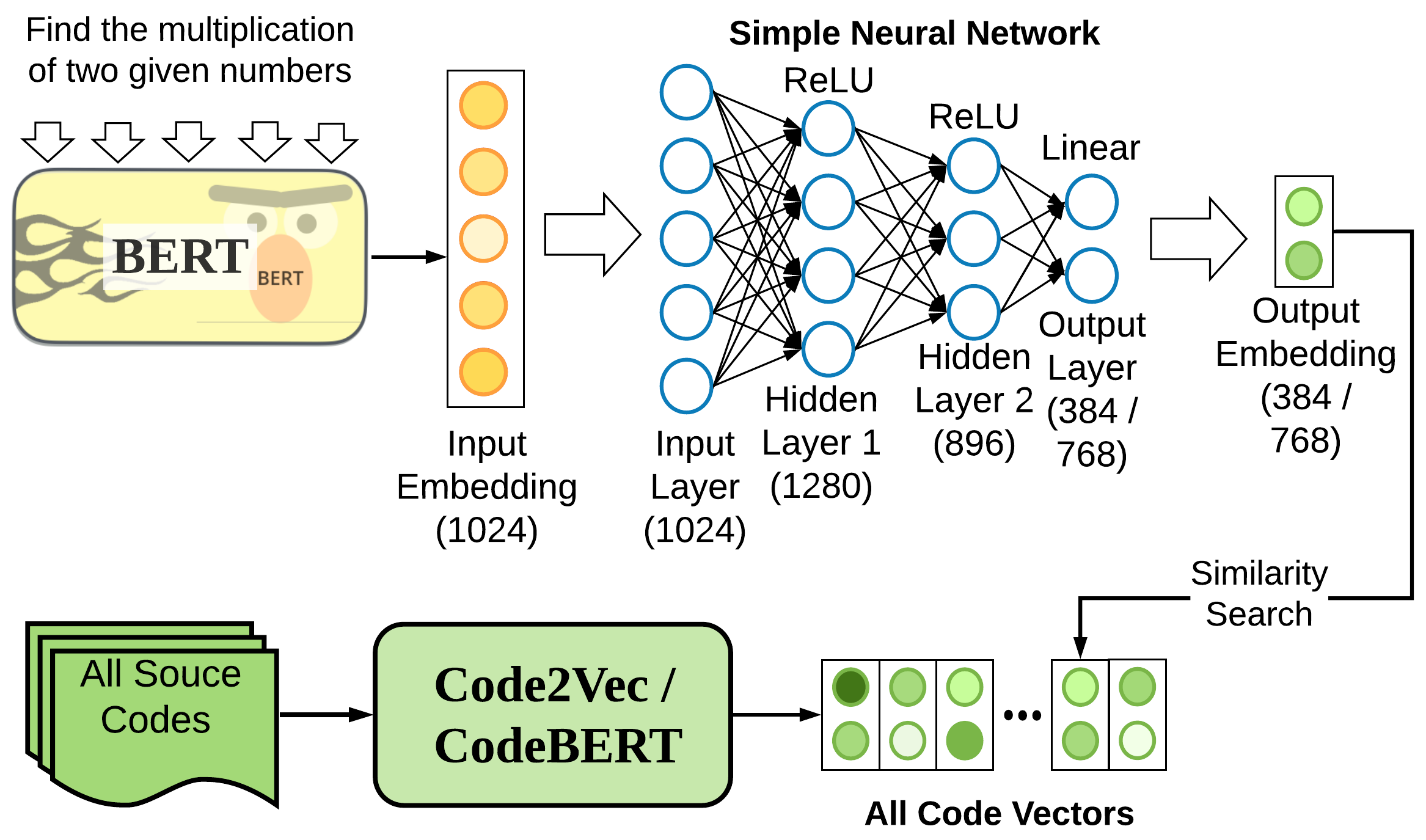}
\caption{Searching network prediction for a new query from all code embeddings in our dataset}
\label{fig:test}
\end{figure}

When the training phase is complete, any new natural language query is, at first, converted into an NL vector. Then, using a trained neural network, we transform the NL vector into a code vector prediction. We then find $n$ code snippets that have code vectors from a given corpus that are closest in terms of L2 distance to the predicted vector, where $n$ is any suitable number. For accelerating this step, we have used FAISS \cite{faiss}. The prediction methodology is presented in Figure \ref{fig:test}.

\subsection{Evaluation Criteria}

%% Describe MRR scoring mechanism
We evaluate the networks' performance with the test set of CodeSearchNet challenge \cite{husain2019codesearchnet} that were unseen during the training phase. In this case, we fix a set for each test sample with 999 distractor datapoints in the same way as the original paper. Then, for each query in this set, we find the closest code vectors and corresponding source codes, and calculate the Mean Reciprocal Rank (MRR) score \cite{mrr-1}, which is used by several related works \cite{unif,feng2020codebert,husain2019codesearchnet,sachdev-ncs}. The Results from this endeavor are described in Section \ref{sec:bert2code-results}.

\vspace*{-1cm}

%% file: Sections/Results.tex
\section{Results}
\label{sec:bert2code-results}
The results obtained from BERT2Code are shown in Table  \ref{tab:bert2code-results}. In our experiment, BERT2Code model with Code2Vec \cite{alon2019code2vec} embeddings have performed better than with CodeBERT \cite{feng2020codebert} embeddings in semantic code search. It is also evident that our model learns the inherent relationship between the embedding spaces as it performs almost 20 times better in the best setting than a random model. 

\begin{table}[h]
\centering
\caption{Semantic code search with BERT2Code}
\label{tab:bert2code-results}
\begin{tabular}{l|c|c|c|c}
\hline
\textbf{Method} & \textbf{Output Layer} & \textbf{Trainable} & \textbf{Epochs} & \textbf{MRR (\%)} \\ 
 & \textbf{Size} & \textbf{Parameters} & & \\ \hline

BERT2Code With & \multirow{2}{*}{384} & \multirow{2}{*}{2,804,224} & \multirow{2}{*}{207} & \multirow{2}{*}{15.478} \\ 
Code2Vec Embeddings & & & & \\ \hline
 BERT2Code With & \multirow{2}{*}{768} & \multirow{2}{*}{3,148,672} & \multirow{2}{*}{170} & \multirow{2}{*}{9.986}  \\ 
CodeBERT Embeddings& & & & \\ \hline
Random Model & - & - & - & 0.74 \\ 
\hline
\end{tabular}
\end{table}

\vspace*{-1cm}

\subsection{Implications of the Results}
In recent years, pretrained embedding models have helped researchers achieve outstanding performances in numerous natural language processing tasks \cite{use-nlembeddings-2,use-nlembeddings-1}. In particular, Cheng and Callison-Burch \cite{balm} has shown that a simple feed-forward neural network can perform considerably well in Neural Machine Translation using BERT pretrained models \cite{devlin2018bert} and achieved a near SOTA performance in the Multi30k English-to-German translation dataset \cite{elliott-etal-2016-multi30k}.

We adopted a similar approach by trying to leverage available state-of-the-art embedding methods for NL queries and code snippets with a lightweight neural network. Our study finds that the model is technologically feasible, can learn the inherent semantic relationship between NL queries and code snippets, and can often find good code suggestions. However, at this moment, it is not the best approach to code search \cite{feng2020codebert,husain2019codesearchnet}. With these findings, we argue that as natural language embeddings have already proven to be readily useful in similar approaches \cite{balm}, in our case, code embeddings can be inferred the bottleneck for our method. 

\subsection{Quality of Code and Sentence Embeddings}

Our simple, and lightweight neural network based code search method largely depends on the embedding models' capability to capture the semantic meanings of source codes and NL queries. We evaluate the code and the sentence embeddings by manually rating queries and source codes based on semantic similarity and then comparing these manual scores with the distances in embedding spaces.

We perform this analysis on the DeepCom dataset \cite{deepcom} to free ourselves from any bias from previous experiments. We generate embeddings with SBERT \cite{reimers2019sbert} and with Code2Vec \cite{alon2019code2vec} (as it performed better than CodeBERT \cite{feng2020codebert}) models for the 505,188 code-NL datapoints from DeepCom. We sample 150 pairs of datapoints that have low Euclidean distance code embedding space and another 150 pairs of datapoints that have low Euclidean distance in the NL embedding space. We then give a similarity score for the two source codes and another similarity score for the NL queries for each of the 300 pairs. The scores are integers ranging between 0 to 10. The scoring is done twice independently by two authors based on the their manual observation. Their ratings for the source codes and the NL queries have 0.8422 and 0.8760 Pearson's correlation \cite{pearsons} respectively. The scores from the two authors are averaged to get our manual semantic similarity score.

\begin{table}
    \centering
    \caption{Comparison between Manual Similarity Scores and the Corresponding Distances in Embedding Spaces in Pearson's Correlation Coefficient and p-value}
    \label{tab:result_similarity_analysis}
    \begin{tabular}{c|c|c|c}
        \hline
        \multicolumn{2}{c|}{\textbf{Measured Variable Pairs}} & \textbf{Correlation} & \textbf{p-value} \\
        \hline
        %NL Similarity by Grader 1 & NL Similarity by Grader 2 & 0.807 & $<10^{-6}$ \\
        %Code Similarity by Grader 1 & Code Similarity by Grader 2 & 0.849 & $<10^{-6}$ \\
        %Manual NL Similarity & Manual Code Similarity & 0.177 & 0.002 \\
        Manual NL Similarity & NL Vectors L2 Distance & -0.772 &  $<10^{-6}$ \\
        Manual Code Similarity & Code Vectors L2 Distance & -0.444 & $<10^{-6}$ \\
        %NL Vectors L2 Distance & Code Vectors L2 Distance & -0.547??? & $<10^{-6}$ \\
        \hline
    \end{tabular}
\end{table}

From Table \ref{tab:result_similarity_analysis}, the code vectors have a lower correlation coefficient with the manual scores than NL vectors. We calculate the \textit{p-value}s for the similarity scores with the null hypotheses that \textit{the manual similarity scores do not correlate with the NL and the code embedding similarities}. Both the \textit{p-value}s are less than $10^{-6}$ which reject the null hypotheses and show that the correlations are statistically significant. This finding supports our assumption that embeddings can be used for learning complex relations between source codes and NL queries. However, the significantly low correlation for the code embeddings similarity indicates that better code embedding methods than Code2Vec \cite{alon2019code2vec} and CodeBERT \cite{feng2020codebert} are necessary to exploit the capability of our method fully.

%% file: Sections/Related_Works.tex
\section{Related Works}
\label{sec:bert2code-related-work}

Our results indicate that better code embeddings are needed to be readily usable in semantic code search. Recently,  Kang et al. \cite{assess-code2vec} has also found that the Code2Vec \cite{alon2019code2vec} embeddings fail to generalize in the tasks of code comments generation, code authorship identification, and code clones detection.

Briem et al. \cite{briem2019using} used distributed representations by Code2Vec \cite{alon2019code2vec} in the task of bug detection. 
Arumugam \cite{arumugam2020semantic_code2vec} trained a Code2Vec bag-of-paths embedding model for the CodeSearchNet Challenge \cite{husain2019codesearchnet}. 

Gu et al. \cite{gu2018deep} introduced CODEnn, which jointly embeds descriptions and source codes into a single high-dimensional vector space. 
Sachdev et al. \cite{sachdev-ncs} proposed Neural Code Search (NCS), a method to extract a sequence of natural language tokens from a code snippet that forms a code document.  
Embedding Unification (UNIF) \cite{unif} is a supervised extension of the NCS. The idea of both these methods is to extract natural language components from the method and identifier names, and use them to match with the natural language query.

The CodeSearchNet Challenge \cite{husain2019codesearchnet} initiated an open challenge for code search, and publicly released a dataset collected from GitHub. Additionally, they released several baseline code search methods. Feng et al. \cite{feng2020codebert} proposed CodeBERT, a BERT \cite{devlin2018bert} model trained with both natural language and source code from the CodeSearchNet dataset \cite{husain2019codesearchnet}, and has shown to significantly outperform the baseline models of the challenge. 

%% file: Sections/Conclusion.tex
\section{Conclusion}
\label{sec:bert2code-conclusion}

In this work, we propose BERT2Code, a simple neural network architecture that performs semantic code search by utilizing pre-trained natural language and code embedding models. We show that our proposed method learns to map natural language embeddings to code embeddings to a reasonable extent. There are multiple findings of our work. Firstly, we show that it is possible to utilize pre-trained embedding models to semantic code search. Secondly, We manually analyze the embedding methods and show that better code embedding methods are needed for better results of our work. Through this work, we would like to draw the attention of the research community to build more generalizable code embedding models. %In future work, we would like to explore other source code analysis tasks, such as code summarization, code generation, code repair, etc utilizing embedding models.